\documentclass[prd,twocolumn,nofootinbib]{revtex4}

\usepackage{amsmath,amssymb,graphicx,subfigure}%\usepackage{amsfonts}

\begin{document}

\preprint{\hepth{0703206}}

\title{Landscape Predictions from Cosmological Vacuum Selection}

\author{Raphael Bousso and I-Sheng Yang\footnote{bousso@lbl.gov, 
jingking@berkeley.edu}}

\affiliation{Department of Physics and Center for Theoretical
Physics \\
University of California, Berkeley, CA 94720, U.S.A. \\
{\em and}\\
Lawrence Berkeley National Laboratory, Berkeley, CA 94720, U.S.A. }

\begin{abstract}%
  In BP models with hundreds of fluxes, we compute the effects
  of cosmological dynamics on the probability distribution of
  landscape vacua.  Starting from generic initial conditions, we find
  that most fluxes are dynamically driven into a different and much
  narrower range of values than expected from landscape statistics
  alone.  Hence, cosmological evolution will access only a tiny
  fraction of the vacua with small cosmological constant.  This leads
  to a host of sharp predictions.  Unlike other approaches to eternal
  inflation, the holographic measure employed here does not lead to
  ``staggering'', an excessive spread of probabilities that would doom
  the string landscape as a solution to the cosmological constant
  problem.
\end{abstract}

\maketitle

\section{Introduction}

The task of making predictions in the landscape of string theory
represents an enormous challenge.  We must survey the metastable vacua
in the theory and find the relative abundance of various low-energy
properties.  But vacua without observers will not be observed.  The
great variability of low energy physics in the string landscape
demands novel methods for capturing such selection effects.

Interposed between these two tasks is a third problem, which will be
the focus of this paper: vacuum selection by cosmological dynamics.
It is not enough for a vacuum to exist in theory.  It will be relevant
only if it can be dynamically realized from generic initial
conditions,  either as regions in spacetime or
branches of a wavefunction.  

The former viewpoint insists on treating spacetime semiclassically on
a global scale.  This has no operational meaning, since no observer
can see beyond his own horizon~\cite{BouFre06a}.  Moreover, it leads
to problematic infinities and ambiguities that have plagued the
definition of measures in eternal inflation; see, e.g.,
Refs.~\cite{LinLin94,GarLin94,GarVil01,GarSch05,EasLim05,Bou06,Bou06b}
for discussions and some recent proposals.  Thus, the global viewpoint
seems poised to suffer the same fate as the ether: a false convenience
increasingly recognized as a burden.  (A similar conclusion was
reached earlier by studying quantum aspects of black
holes~\cite{SusTho93,Pre92}, and it seems natural indeed that it
should extend to cosmology.)

One of us recently proposed a ``holographic'' measure~\cite{Bou06}
that refers only to a single causally connected region, or causal
diamond.  In Ref.~\cite{BouHar07}, it was shown that this measure not
only overcomes problems that plagued anthropic predictions of the
cosmological constant, but is able to maintain this success when
specific anthropic conditions are traded for the much weaker
assumption that observers require merely free energy.  

Here we apply a different aspect of the proposal of Ref.~\cite{Bou06}.
We focus not on anthropic selection, but on cosmological selection.
We shall find that dynamical effects can significantly suppress or
enhance the probability of observing a given vacuum, and they can
interfere with anthropic selection in some models.  However, unlike
another recently proposed measure~\cite{GarSch05}, the holographic
measure does not lead to excessively uneven probability distributions
that would render the string landscape ineffective at solving the
cosmological constant problem.

We will follow Ref.~\cite{BP} in modelling the string landscape as
arising from a large number of possible four-form flux configurations.
We mainly consider a model with $250$ fluxes, with fixed charges of
order $1/30$.  This yields a large number of metastable vacua, of
which $10^{121}$ have vacuum energy comparable to the observed value.
Of these, however, we find that only $10^{80}$ are accessed by a
typical worldline, starting from generic initial conditions.

The selected ensemble is characterized by 250 probability
distributions over the integers, one for each flux.  The distributions
differ---in many cases, drastically---from the distributions one would
have obtained by simply restricting the landscape to vacua with small
cosmological constant.  Thus, cosmological selection leads to
thousands of distinct predictions.  Many correspond to probabilities
that are so close to 0 or 1 that even a single conflicting observation
would rule out the model.

The specific predictions we obtain apply only to the toy model we
study, but they do allow us to draw more general lessons.  Most
importantly, our results demonstrate that cosmological dynamics can be
a powerful constraint.  It reduces the effective size of the landscape
drastically, leading to a large number of strong predictions.  Another
general lesson is that fast decays happen first.  In natural models,
it takes hundreds or thousands of tunneling events to get from a
generic initial vacuum to a vacuum with observers.  Any metastable
internal configuration that gives rise to a relatively fast decay
channel but is unlikely to arise late in the decay chain will not be
observed.

In the real string landscape, the selection effects will likely be as
strong as in the BP models studied here, but they may be harder to
characterize in terms of a set of nearly independent flux
distributions.  For example, charges will not be independent of
fluxes, and their dynamical effects can become intertwined.  

The specific flux configuration associated with our own vacuum is
unlikely to be observed directly in the near future.  However, fluxes
are related to coupling constants and other low-energy parameters.  It
will be important to understand such correlations in realistic and
representative sets of string compactifications.  Then it will be
possible to understand the impact of cosmological selection effects on
the distribution of low-energy parameters.

\paragraph{Outline}
Sec.~\ref{sec-mm} includes brief descriptions of the BP model, and of
the holographic cosmological probability measure we use to compute
cosmological selection effects.  We focus on the question of how this
measure can be effectively applied to a model with ten to the hundreds
of vacua.  This would seem to require inverting a matrix of the
corresponding rank.  However, we develop statistical methods that are
both adequate and very efficient.  (Details of the techniques are
given in Appendices~\ref{sec-independent} and \ref{sec-unbiased}.) We
also discuss our choice of initial conditions, and we show that our
results will be the same for any initial probability distribution that
depends only on $\lambda$ and does not strongly favor small or
negative values of $\lambda$.

In Sec.~\ref{sec-results}, we present our results.  For a
representative model, we contrast the distribution of fluxes, and the
number of vacua with small cosmological constant, before and after
cosmological selection effects are taken into account.  We point out
specific predictions that can be made with high confidence, focussing
on features that would have been either absent or strikingly different
if cosmological dynamics had been neglected.  We consider other models
to illustrate that the strength of cosmological selection effects
varies and that, in special cases, a model that naively contains
enough vacua may fail after cosmological selection is accounted for.

In Sec.~\ref{sec-discussion}, we try to give a qualitative explanation
of the features seen in the cosmological flux distribution.  We
identify them as imprints of the dynamics at different stages in the
decay chain, when different types of transitions dominate.  

In Sec.~\ref{sec-discussion2}, we compare our results to those
obtained by Schwartz-Perlov and Vilenkin \cite{SchVil06} from a
different proposal for the probability measure~\cite{GarSch05}.  In
toy-BP-models with fewer vacua, the alternative measure leads to a
``staggered'' probability distribution, which would leave the string
landscape effectively underpopulated and unable to solve the
cosmological constant problem.  The measure of Ref.~\cite{GarSch05} is
difficult to apply explicitly in models with a realistic number of
vacua, and to extend beyond first order in perturbation theory.
However, we show in Appendix~\ref{sec-mimic} that it is equivalent to
the holographic measure with particular (and from our point of view,
highly unnatural) initial conditions.  With the help of this trick, we
are able to argue that the staggering problem is likely to persist at
higher orders when the measure of Ref.~\cite{GarSch05} is applied to
large BP models.

\section{Model and measure}
\label{sec-mm}

\subsection{Probability measure, course graining, and Monte Carlo
  chains}
\label{sec-grain}

Consider a worldline starting in some metastable initial vacuum.  This
vacuum will eventually decay, giving way to a new vacuum.  The cascade
continues until the worldline enters a terminal vacuum, which does not
decay further.  According to the proposal of Ref.~\cite{Bou06}, the
probability $P_A$ assigned to vacuum $A$ is the expected number of
times the worldline will enter vacuum $A$ on its way down to a
terminal vacuum.\footnote{In Ref.~\cite{Bou06}, $A, B, \ldots$
  represented specific vacua.  Here they are indices running over all
  vacua.  In Appendix~\ref{sec-mimic} we use indices $a,b,\ldots$ to
  run over metastable vacua only.  Indices $i,j,\ldots$ are reserved
  for fluxes.}  For a more general initial probability distribution
$P_A^{(0)}$, one sums over the result obtained from each initial
vacuum, weighted by $P_A^{(0)}$.

The probabilities $P_A$ form an ${\cal N}$-dimensional vector, where
${\cal N}$ is the number of vacua in the landscape.  It satisfies the
matrix equation~\cite{Bou06}
\begin{equation}
(1-\eta) \mathbf P = \eta \mathbf P^{(0)}~.
\label{eq-matrix}
\end{equation}
where $\eta_{AB}$ is the relative decay rate, or branching ratio, from
vacuum $B$ into vacuum $A$.  It can be obtained from the decay rate
per unit time, $\kappa_{AB}$, by normalizing each column of
$\kappa_{AB}$,
\begin{equation}
\eta_{AB} = \frac{\kappa_{AB}}{\sum_C\kappa_{CB}}~,
\label{eq-rel}
\end{equation}
except for columns corresponding to terminal vacua (which vanish both
for $\kappa$ and for $\eta$).

In a large landscape, one expects that both the initial distribution,
$P_A^{(0)}$, and the distribution resulting from dynamical selection,
$P_A$, will have support over an enormous number of vacua.  Then it is
impractical to solve Eq.~(\ref{eq-matrix}) in detail.  But
fortunately, the probability of a particular vacuum is no more
interesting than that of a specific quantum state in a macroscopic
system.  A more natural set of questions is the following: How large
is the subset of vacua selected by cosmological dynamics, and how can
it be characterized?  

In fact, we will ask a more restrictive set of questions.  Let us
suppose that vacua with $\lambda\gg\lambda_0$ contain no observers,
where
\begin{equation}
\lambda_0 \approx 7.9\times 10^{-121}
\label{eq-cc}
\end{equation}
is the observed cosmological constant.  Then we may restrict our
attention to vacua with cosmological constant $\lambda$ of order the
observed value:
\begin{equation}
0<\lambda\lesssim\lambda_0~.
\label{eq-weinberg}
\end{equation}
(For simplicity, we exclude negative values of $\lambda$; this will
make our formulas simpler without changing our results qualitatively.
For the same reason, we will take the upper bound to be sharp.)  We
will consider only models which contain a large number of such vacua,
and we will be interested in the restriction of the probability
distribution $P_A$ to this subset of vacua.  Thus, we ask:
\begin{itemize} 
\item[(1)] How large is the set of vacua which (a) satisfy
Eq.~(\ref{eq-weinberg}) and (b) are selected by cosmological dynamics?  
\item[(2)] What characteristics distinguish typical vacua in the above
  set from typical vacua which merely satisfy (a)?
\end{itemize}

Instead of solving Eq.~(\ref{eq-matrix}), we will approach these
questions by a Monte Carlo simulation.  We select an unbiased sample
of $N$ initial vacua according to a generic initial distribution
$P_A^{(0)}$.  For each initial vacuum, we simulate a decay chain: At
every step, a new vacuum $B$ is selected with a probability given by
the branching ratios, $\eta_{AB}$, where $B$ is the old vacuum.  The
chain will eventually reach a terminal vacuum, i.e., a vacuum with
negative cosmological constant.  We are interested in the penultimate
vacuum, which still has positive vacuum energy.  For large enough $N$,
the penultimate vacua obtained from the Monte Carlo simulation become
a representative sample of the ensemble of vacua selected from the
landscape by cosmological dynamics.

\subsection{BP model and decay rates}
\label{sec-setup}

\begin{figure*}
\begin{center}
\includegraphics[scale = .6]{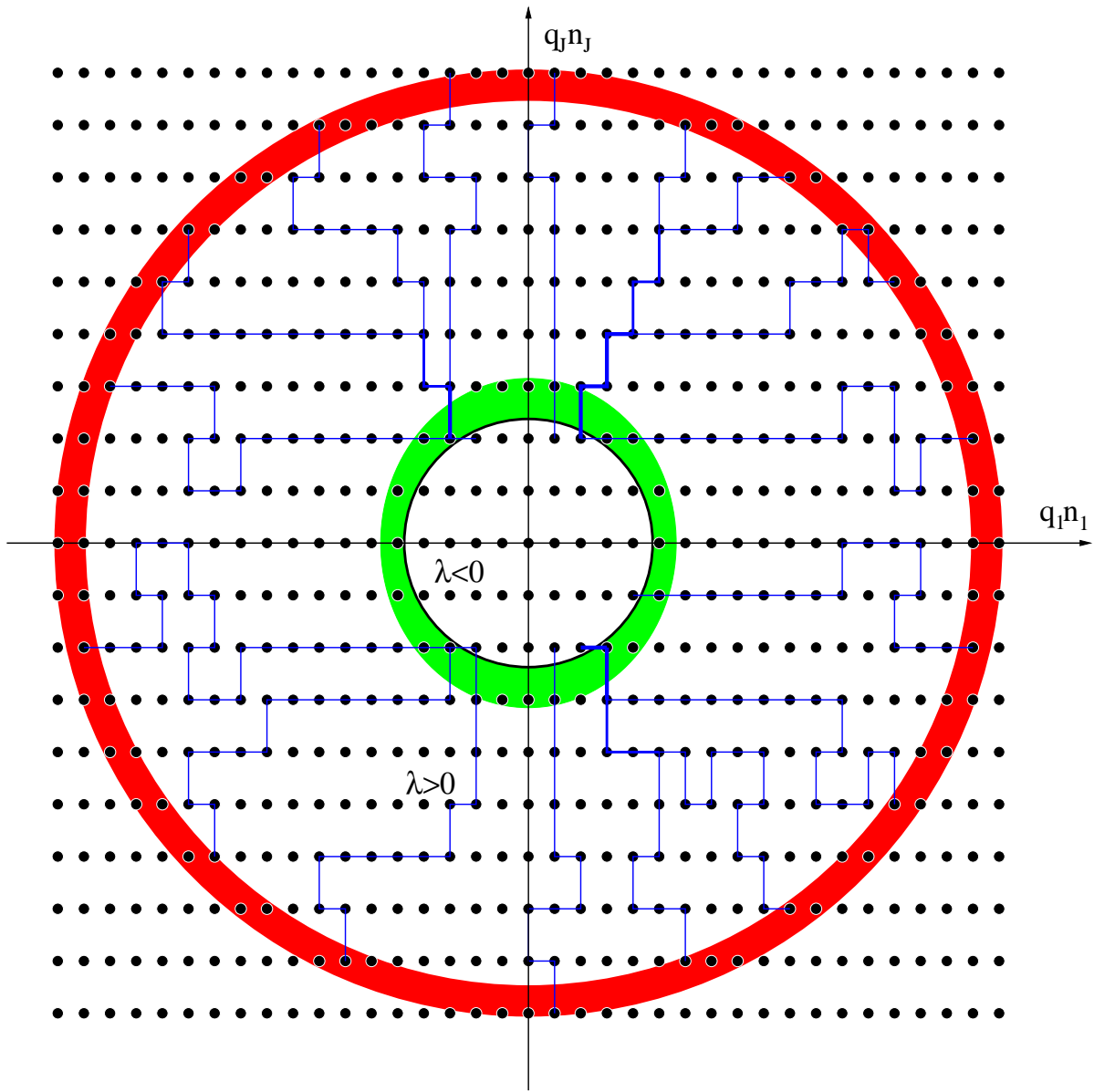}
\end{center}
\caption{A two-dimensional slice through the $J$-dimensional
  flux grid.  Surfaces of constant $\lambda$ are $J-1$ dimensional
  spheres.  Each dot represents a vacuum.  Starting from initial vacua
  with $\lambda\approx 1$ (outermost shell), we simulate decay chains
  (blue lines) through the landscape, which terminate when a vacuum
  with $\lambda\leq 0$ is reached.  Among the ``penultimate'' vacua
  with relatively small positive $\lambda$ (green/shaded region), only
  a small fraction is actually accessed by the decay chains, leading
  to a host of predictions.  Vacua with $\lambda$ of order the tiny
  observed value lie in a much thinner shell (schematically shown as a
  black circle).  A model can be ruled out if the selection effects
  render this shell inaccessible.}
\label{fig-para}
\end{figure*}

We will follow Ref.~\cite{BP} in modeling the string landscape as a
set of flux vacua, neglecting the backreaction of fluxes and moduli
stabilization.  The geometry is of the form $M_4\times X$, where $X$
is a compact 6 or 7-dimensional manifold, and $M_4$ is the macroscopic
four-dimensional spacetime.  A $p$-brane wrapped on a $p-2$ cycle in
$X$ acts as a membrane in $M_4$.  The number of cycles, $J$, and thus
the number of membrane species, can range into the hundreds.  Each
species sources a four-form field strength in $M_4$.  The four-form
flux is quantized, so each vacuum is fully specified by the set
$\{n_i\}$, $i=1,\ldots, J$, of integer flux numbers.  The vacua form a
$J$ dimensional grid in flux space (Fig.~\ref{fig-para}).

The cosmological constant\footnote{We use units in which $8\pi
  G=c=\hbar=1$.}  can be written as
\begin{equation}
\lambda=\lambda_{\rm bare}+\frac{1}{2}\sum^J_{i=1}n_i^2q_i^2~,
\label{eq-BP}
\end{equation}
Here, $q_i$ is the charge of the $i$-th membrane.  The second term is
contributed by the $J$ species of four-form flux.  $\lambda_{\rm
  bare}$ subsumes all other contributions, in particular those from
vacuum loops, and will be assumed to be naturally large and negative:
\begin{equation}
-\lambda_{\rm bare}=O(1)~.
\end{equation}

If each flux can take several different integer values, then there
will be an enormous number of vacua.  Some of these vacua will
accidentally have a very small cosmological constant, even though no
small scale is introduced into the model.  If the model contains vacua
with a cosmological constant on the order of the observed value in our
universe, $\lambda_0$, then the cosmological constant problem can be
solved by anthropic selection effects~\cite{Wei87,Bou06,BouHar07}.

A more detailed model of the string landscape should include the
backreaction of fluxes on the geometry (see Ref.~\cite{DouKac06} for a
review).  These effects are still hard to control, especially if we
are interested in the cosmology of a large, representative part of the
landscape.  In particular, both the charges $q_i$ and the bare
cosmological constant would depend on the set $\{n_i\}$.  However,
this is unlikely to affect a key feature of the BP model: that the
vacua form a discretuum with tiny effective spacing of the
cosmological constant.

In general, a metastable vacuum can decay to any other vacuum.
However, a typical decay will change one flux by one unit.  Tunneling
events that change more than one flux, or change a flux by more than
one unit, can be regarded as a composite event.  They will be much
more suppressed than one of their constituent events and can thus be
neglected.  This simplifies our task, reducing the number of possible
decays at each step to $2J$.  The only question is which flux will
change, and in which direction.

The relative decay rates are given by Eq.~(\ref{eq-rel}), with
\begin{equation}
\kappa_{ij} =A_j e^{-B_{ij}}~.
\end{equation}
The constant $A_j$ depends only on the host vacuum and drops out in
the branching ratios.  The instanton action for decreasing the
magnitude of the $i$-th flux, $n_i$, by one unit is given
by~\cite{BroTei87}
\begin{eqnarray}
  B(|n_i|\to |n_i|-1) &=& B_{\rm flat}(|n_i|\to |n_i|-1)\, r(x,y)
  \nonumber \\ 
  B_{\rm flat}(|n_i|\to |n_i|-1) &=& 
  \frac{27\pi^2}{8}\frac{1}{\left(|n_i|-\frac{1}{2}\right)^3q_i^2}  
  \nonumber \\
  r(x,y)&=&\frac{2[(1+xy)-\sqrt{1+2xy+x^2}]}
  {x^2(y^2-1)\sqrt{1+2xy+x^2}} \label{eq-B} \nonumber \\ 
  x&=&\frac{3}{8\left(|n_i|-\frac{1}{2}\right)} \nonumber \\
  y&=&\frac{2\lambda}{q_i^2\left(|n_i|-\frac{1}{2}\right)}-1~,
\end{eqnarray}
where $\lambda$ is the cosmological constant in the first vacuum.
Upward tunneling is further suppressed by the relative entropy of the
two vacua:
\begin{eqnarray}
 & B(|n_i|\to |n_i|+1)= & \\ & B(|n_i|+1\to |n_i|)
  \exp\left(\frac{24\pi^2}{\lambda+\left(|n_i|+\frac{1}{2}\right) q_i^2}
    -\frac{24\pi^2}{\lambda}\right) & ~. \nonumber
\label{eq-upward}
\end{eqnarray}
For fluxes with $n_i=0$ it is important to take into account that
$0\to 1$ and $0\to -1$ are two distinct decay channels that contribute
equal amounts to the total decay rate entering Eq.~(\ref{eq-rel}).

\subsection{Quantifying selection effects}
\label{sec-quantify}

Given the $N$ vacua selected by our Monte Carlo simulation, we can
answer the two questions posed above.  First, let us focus on how the
selected vacua can be characterized and distinguished from an unbiased
ensemble of vacua with small cosmological constant.

In our toy model, the only variables we have available are the fluxes,
$n_i$.  A list of $J$ fluxes specifies a unique vacuum, but a list of
$J$ probability distributions (one for each flux) defines an ensemble
of vacua.  Larger ensembles could be defined by throwing away more
information, e.g., by specifying only a list of $J$ expectation values
$\langle n_i\rangle$.  However, we are interested in extracting as
much information as possible from our simulation.  Hence, we will
consider the $J$ discrete distributions
\begin{eqnarray}
  &p_i(n) \equiv \mbox{probability that the}~i\mbox{-th flux,}\, 
  n_i,&\nonumber\\ & \mbox{has integer value}~n~. &
\end{eqnarray}
Selection effects can be demonstrated by comparing the $J$
distributions obtained from the selected vacua, with the $J$
distributions obtained from an unbiased sample of vacua with a similar
range of vacuum energy.

Next, let us turn to estimating the total number of selected vacua
from the small sample obtained by the simulation, and comparing this
to the total number of vacua in a similar range of vacuum energy.
This task is slightly more subtle. 

Associated with each probability distribution $p_i(n)$ is a Shannon
entropy,
\begin{equation}
s_i=-\sum_{n=-\infty}^\infty p_i(n) \log p_i(n)~.
\end{equation}
We shall treat the $n_i$ as independent random variables.  This
assumption is justified, and small corrections are given, in
Appendix~\ref{sec-independent}.  Then the total entropy is
\begin{equation}
S=\sum_{i=1}^J s_i= -\sum_{n,\,i} p_i(n) \log p_i(n)~,
\label{eq-ent}
\end{equation}
and the corresponding number of ``states'' (i.e., vacua) in the
ensemble is
\begin{equation}
{\cal N} = \exp S\, .
\label{eq-exp}
\end{equation}
We will be interested in computing this number not only from the
selected sample of $N$ vacua, but also from an unbiased sample of
vacua with a similar range of vacuum energy.  Their fraction,
\begin{equation}
  \frac{\cal N}{{\cal N}^{\rm unselected}}
\end{equation}
quantifies the degree to which cosmological selection thins out the
landscape.

We are especially interested in how many distinct values of the
cosmological constant, $\lambda$, are contained in the spectrum of
vacua, before and after cosmological selection.  By Eq.~(\ref{eq-BP}),
changing the sign of a flux, $n_i\to -n_i$, leaves $\lambda$
invariant.  To account for this degeneracy, let us consider a new set
of distributions in which vacua are treated as identical if they
differ only by the signs of fluxes:
\begin{equation}
\tilde p_i(n) = \left\{ 
\begin{array}{ll}
p_i(0)\, , & n=0\,;\\
p_i(-n)+p_i(n)\, , &n>0\, .
\end{array}\right.
\label{eq-ptilde}
\end{equation}
From this set of $\tilde p_i$, we can compute the number $\tilde {\cal
  N}$ of different values of the cosmological constant, by the
analogues of Eqs.~(\ref{eq-ent}) and (\ref{eq-exp}).  

As we remarked above, we assume an initial probability distribution
that depends only on $\lambda$.  This ensures that $p_i(-n)=p_i(n)$
initially (up to finite-sample effects), and by since the decay rates
depend only on $|n_i|$, the condition is preserved by dynamics.
Hence,
\begin{equation}
p_i(n)= p_i(-n) = \left\{ 
\begin{array}{ll}
\tilde p_i(0)\, , & n=0\,;\\
\frac{1}{2}\tilde p_i(|n|)\, , &n \neq 0\, .
\end{array}\right.
\end{equation}
For this reason, we only display $\tilde p_i(n)$ in all figures below.

The $N$ vacua in our sample will be distributed over a much wider
interval of $\lambda$ than that of Eq.~(\ref{eq-weinberg}); very
likely, none of them will have $\lambda\sim\lambda_0$.  However, the
distribution of $\lambda$ is smooth on scales much larger than
$\lambda_0$, so that the total number of dynamically selected values
of $\lambda$ satisfying Eq.~(\ref{eq-weinberg}), $\tilde{\cal N}_{\rm
  W}$, can be easily estimated:
\begin{equation}
\tilde{\cal N}_{\lambda_0}=\lambda_0 \left.
\frac{d\tilde {\cal N}}{d\lambda}\right|_{\lambda=0}
\approx
\lambda_0 \frac{\tilde{\cal N}}{N} \left.
\frac{dN}{d\lambda}\right|_{\lambda=0}~,
\label{eq-nw}
\end{equation}
where the distribution $dN/d\lambda$ can be obtained from our Monte
Carlo simulation after suitable binning, and, from Eqs.~(\ref{eq-ent})
and (\ref{eq-exp}),
\begin{equation}
  \tilde {\cal N}=\exp[-\sum \tilde p_i(n)\log\tilde p_i(n)]~.
  \label{eq-tilden}
\end{equation}

In Appendix~\ref{sec-independent} we note that the fluxes are only
approximately, but not completely independent random variables.  We
argue that an appropriate correction can be implemented by generating
a new sample of $N'$ vacua from the probability distributions $\tilde
p_i(n)$ obtained from the penultimate vacua on the Monte Carlo chain,
and replacing $N$ by $N'$ in Eq.~(\ref{eq-nw}).

In Appendix~\ref{sec-unbiased} we explain how to estimate the number
$\tilde {\cal N}_{\lambda_0}^{\rm unselected}$ of distinct values
$\lambda\sim\lambda_0$ in the model (before cosmological selection).
An approximate formula was given in Ref.~\cite{BP} but we propose a
more precise method that better accounts for degeneracies of vacua.

\subsection{Initial conditions}
\label{sec-initial}

There is no universally accepted theory of initial conditions, and it
is not our purpose to propose such a theory.  Given initial
conditions, we are interested in the effects of the cosmological
dynamics on the selection of vacua with small cosmological constant.
For definiteness, we will assume an initial probability distribution
that depends only on $\lambda$ and favors large values of the vacuum
energy.

This type of initial distribution would follow, for example, by
assuming that all vacua are equally likely.  The number of vacua
with cosmological constant less than $\lambda$ grows very rapidly,
like $(\lambda-\lambda_{\rm bare})^{J/2}$.  Hence, the overwhelming
majority of vacua have large positive values of $\lambda$.

An even stronger preference for large initial $\lambda$ follows if we
adopt the tunneling proposal~\cite{Vil86,Lin84b}:
\begin{equation}
P_i^{(0)}\propto \left\{ 
\begin{array}{ll}
  e^{-24\pi^2/\lambda}\, , & \lambda>0\,;\\
  0\, , &\lambda\leq 0\, .
\end{array}\right.
\label{eq-tunneling}
\end{equation}
This assigns negligible probability to vacua with small cosmological
constant.  Vacua with a large positive vacuum energy are then favored
not only because of their greater number, but because of their greater
individual probability.

It makes little difference which of these specific proposals we adopt.
The above distributions diverge at large values of $\lambda$ and must
be cut off, so most initial vacua will be at the cutoff.  We will
choose a cutoff $\lambda\lesssim 1$.  It seems reasonable to ignore
vacua with $\lambda\gtrsim 1$ since semiclassical gravity cannot be
valid in this regime.  One signature of this breakdown is that action
of instantons mediating their decay becomes of order unity or less.
This means that the vacua cannot be treated as distinct, metastable
states.

\begin{figure}
\begin{center}
\includegraphics[scale = .8]{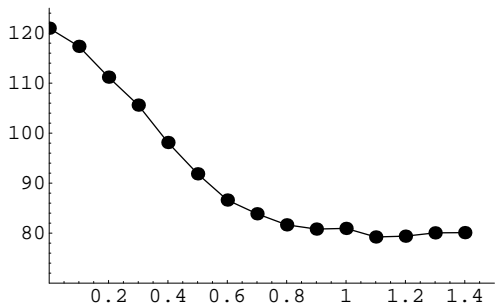}
\end{center}
\caption{The number of selected vacua with $\lambda\sim\lambda_0$,
  $\log{\cal N}_{\lambda_0}$, obtained from different initial
  ensembles starting at different values of $\langle\lambda\rangle$,
  in Model 1.  For initial values $\langle\lambda\rangle\gtrsim 0.8$,
  the number of selected vacua stops shrinking.  This indicates that
  at such high values, selection effects are negligible.  Therefore,
  the details of the initial conditions are irrelevant as long as
  large values of $\lambda$ are preferred, as explained in the text.}
\label{fig-asymptote}
\end{figure}

Fortunately, our results show little sensitivity to the precise choice
of the cutoff.  We have run sets of decay chains starting from
different initial values of $\lambda$.  As the initial value is
increased, the selection effects at first become more important.  This
is not surprising, since a larger initial value implies a longer decay
chain, and hence, more opportunities for preferred decay channels to
imprint directional effects on the probability distribution.
Eventually, however, the distribution shows asymptotic behavior,
changing little as the initial $\lambda$ is further increased (see
Fig.~\ref{fig-asymptote}; in the model here, this occurs near
$\lambda=1$).  This is also reasonable: at large cosmological
constant, all decay channels become unsuppressed.  Directional
selection effects appear to set in only when $\lambda$ becomes
sufficiently small.  Therefore, the precise cutoff, and thus the
choice of initial cosmological constant, do not affect our results.

This clearly alleviates the dependence of the local probability
measure on initial conditions.  However, it does not eliminate this
dependence entirely.  For example, if the initial probability
distribution was concentrated on a particular vacuum (rather than
dependent only on $\lambda$), then the selected vacua would
generically show a significant imprint of this origin.  This is as it
should be.  If we reject the global viewpoint, there is no reason to
expect that all initial conditions must be washed out.

\section{Results}
\label{sec-results}

We will first present results for a model with $J=250$, $\lambda_{\rm
  bare}=-0.5$, and $q_i = 0.01494+0.03\frac{i}{J}$ (``Model
1'').\footnote{Strictly, the charges should be incommensurate in order
  to avoid additional degeneracies.  Since this could be implemented
  by extremely small modifications of the $q_i$, we ignore such
  degeneracies here.}  The model contains ${\cal N}_{\lambda_0}^{\rm
  unselected}\approx 10^{121}$ vacua in the interval $0<\lambda\leq
\lambda_0$, with $\tilde{\cal N}_{\lambda_0}^{\rm unselected}\approx
10^{60}$ distinct values of $\lambda$.  We considered initial
distributions centered on $\langle\lambda\rangle=0.1, 0.2, \ldots,
1.4$.  For each initial distribution (for example, the red/dark shell
in Fig.~\ref{fig-para}), we ran $1000$ Monte Carlo decay chains
selecting $1000$ vacua with small positive cosmological constant (in
the green/light shell in Fig.~\ref{fig-para}).
\begin{figure}
\begin{center}
\includegraphics[scale = .75]{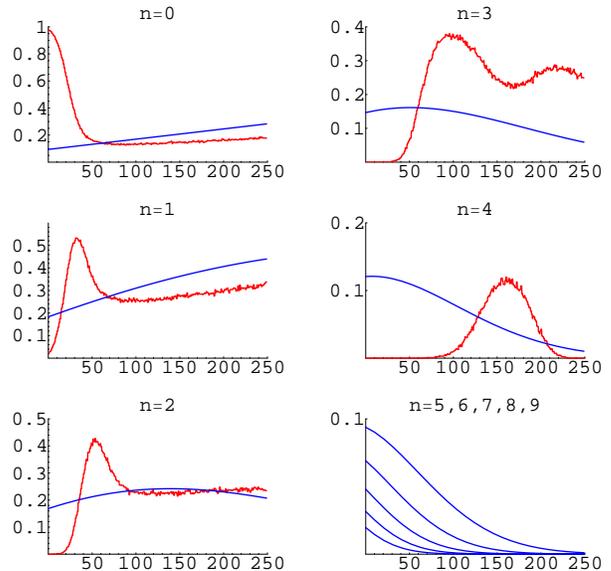}
\end{center}
\caption{The red, wiggly curves show the probability that the $i$-th
  flux is $n$, $\tilde p_i(n)$, in the penultimate vacua of $N=7000$
  decay chains starting from $\lambda\gtrsim 0.8$ in Model 1, as a
  function of $i$.  For comparison, the blue, smooth curves show the
  probabilities that would have been obtained without cosmological
  selection, just from restricting to vacua with small cosmological
  constant.  The differences are immediately apparent and quite
  drastic.  For example, no fluxes greater than 4 survive the
  selection process; without selection, there would be many such
  fluxes.  The fluxes associated with small charges (small $i$) are
  anomalously low after selection.}
\label{fig-pnis}
\end{figure} 

The probability distribution of fluxes characterizing the selected
vacua becomes independent of the initial distribution for initial
values of $\lambda\gtrsim 0.8$ (see Fig.~\ref{fig-asymptote}).  This
asymptotic distribution is what we are after: it is the one picked out
by the initial conditions chosen in Sec.~\ref{sec-initial}.  To
improve the statistical quality of our final data, we therefore merged
the distributions obtained from initial $\langle\lambda\rangle=0.8, 0.9,
\ldots, 1.4$, effectively obtaining $N=7000$ runs.

\begin{figure*}
\begin{center}
\includegraphics[scale = .9]{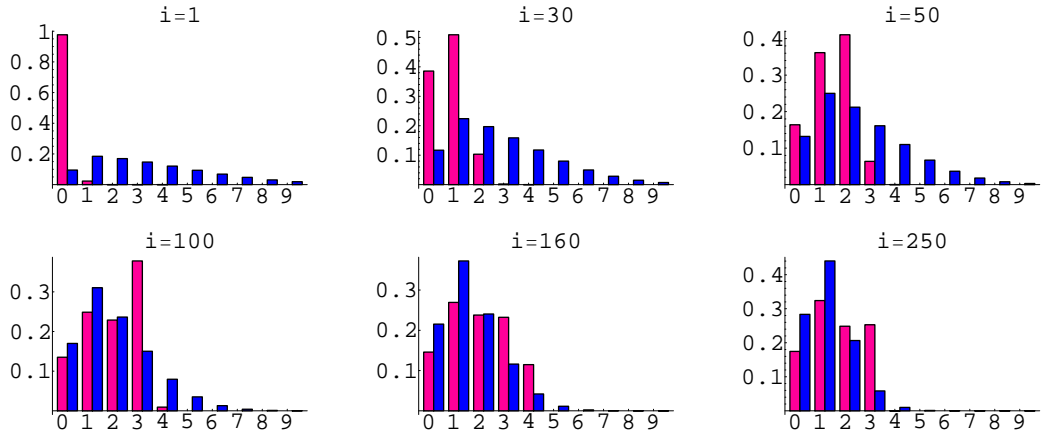}
\end{center}
\caption{Specific examples of probability distributions $\tilde
  p_i(n)$, plotted against $n$ for 6 of the 250 fluxes in Model 1.
  The cosmological dynamics drives most fluxes into a narrower range
  (red/light shaded bars) than the distribution obtained from
  landscape statistics alone (blue/dark bars).}
\label{fig-pn}
\end{figure*}

The resulting probability distribution of fluxes characterizing the
selected vacua, $\tilde p_i(n)$, is shown as a function of $i$ in
Fig.~\ref{fig-pnis}, and as a function of $n$ in Fig.~\ref{fig-pn}.  For
comparison, we show the probabilities that would have been obtained
from a random sample of vacua with comparable cosmological constant,
without cosmological selection.

The overall strength of the selection effects (our first question in
Sec.~\ref{sec-grain}) can be quantified by the ratio of selected vacua
to the total number of vacua with distinct cosmological constant in
the interval $0<\lambda<\lambda_0$ (the thin black shell in
Fig.~\ref{fig-para}).  We find from Eq.~(\ref{eq-nw}) that ${\cal
  N}_{\lambda_0}\approx 10^{80}$, so
\begin{equation}
\frac{{\cal N}_{\lambda_0}}{{\cal N}_{\lambda_0}^{\rm unselected}}
\approx 10^{-41}~.
\end{equation}
In other words, only one out of every $10^{41}$ vacua with small
cosmological constant is actually accessed by cosmological selection.

After taking into account degeneracies, there remain effectively
$\tilde{\cal N}_{\lambda_0}\approx 10^{22}$ distinct values of $\lambda$ that
are accessed by cosmological selection.  Hence, the cosmological
constant problem can still be solved in this model, albeit with a much
reduced discretuum density.

In order to answer the second question posed in Sec.~\ref{sec-grain},
let us inspect the probability distributions $\tilde p_i(n)$ in more
detail, and look for characteristics that distinguish the selected
vacua from the full set of vacua with small cosmological constant.  In
principle, every one of the $J$ fluxes will have a different
probability distribution, leading to $O(J)$ predictions.  For example,
the first flux $n_1$ vanishes with $95\%$ probability.  Before
selection, by constrast, it would have been unlikely to do so ($10\%$;
see the first panel in Fig.~\ref{fig-pn}).

Instead of going through each flux, it is more illuminating to
identify general trends, paying special attention to features that
would have been extremely unlikely in the unselected sample.  Quite
clearly, small charges tend to have smaller fluxes.  Interestingly,
this is precisely the opposite of what would be naively expected
without cosmological dynamics.

Let us define $i_{\rm max}(n)$ to be the largest $i$ such that there
is less than $1\%$ chance to find one or more $n_j>n$ in the range
$1\leq j\leq i$.  For the selected vacua, we have $i_{\rm max}(1)=14$,
$i_{\rm max}(2)=31$ and $i_{\rm max}(3)=77$.  Without cosmological
selection, by contrast, it is practically certain that one or more
fluxes in the above ranges will exceed $n$.

Strictly speaking, we cannot resolve probabilities less than $N^{-1}$,
the limit from the number of Monte Carlo runs.  However, one would
expect to obtain much stronger predictions by involving analytic
arguments.  For example, the $p_i(4)$ graph shows that for each
$i<34$, the probability of finding $n_i=4$ is ``zero'', i.e., less
than $N^{-1}$.  Dynamically, this must have resulted from a strong
preference for such fluxes to decay---a preference that must have been
present over more than $34$ steps in the decay chain.  For example, if
we assume that $n_1=4$ at some point in the decay chain, it must be
true that long before the end of the decay chain, the probability for
it to remain untouched must become very small.  But then it will be
small for each of the remaining steps.  Hence, we expect that $p_1(4)$
is not just less than $N^{-1}$, but in fact exponentially small.

We have studied a range of other model parameters to ensure that our
results are not peculiar features of this model.  For example, in
Model 2 with $J=200$, $\lambda_{\rm bare}=-0.6$, and
$q_i=0.01494+0.03\frac{i}{J}$, we find $\frac{{\cal
    N}_{\lambda_0}}{{\cal N}_{\lambda_0}^{\rm unselected}} \approx
10^{-23}$; the selection effect is smaller but still considerable.
The number of different values of $\lambda\sim\lambda_0$ is
$\tilde{\cal N}_{\lambda_0}^{\rm unselected}\approx 10^{39}$ before
cosmological selection and $\tilde{\cal N}_{\lambda_0}\approx 10^{16}$
after selection.  Thus, as in the previous model, cosmological
selection effects yield predictions but keep the model viable.

Of course, there will also be models which are ruled out by
cosmological selection effects.  The simplest way to find such models
is to reduce the discretuum density before selection, for example by
reducing the number $J$ of fluxes or decreasing $|\lambda_{\rm
  bare}|$.  If the model contains only a few vacua with
$\lambda\sim\lambda_0$ to start with, chances are that there will be
none left after cosmological selection effects are taken into account.
For example, consider Model 3 with $q_i=0.01494+0.03\frac{i}{J}$,
$J=200$ and $\lambda_{\rm bare}=-0.4$.  The discretuum density per
$\lambda_0$ is $\tilde{\cal N}_{\lambda_0}^{\rm unselected} \approx
10^{24}$ before selection but $\tilde{\cal N}_{\lambda_0}\approx
10^{-12}$ after selection.  In other words, not even a single vacuum
with $\lambda\sim\lambda_0$ can be accessed.  This model is ruled out
by cosmological selection.

But the strength of selection effects can also vary.  For example,
Model 4 with $q_i=0.01494+0.03\frac{i}{J}$, $J=150$, $\lambda_{\rm
  bare}=-1.0$ has the same $\tilde{\cal N}_{\lambda_0}^{\rm
  unselected}\approx 10^{24}$ as Model 3.  But after cosmological
selection, the discretuum density per $\lambda_0$ is $\tilde{\cal
  N}_{\lambda_0}=10^{13}$, still larger than $1$, so the model remains
viable.

Finally, we should mention that there can also be models in which
cosmological selection effects are unimportant.  For example, for very
small charges $q_i\ll 1$, the typical fluxes $n_i$ will be large, and
their standard deviations $\delta n_i$ in the interval
$0<\lambda\leq\lambda_0$ can be much larger than the typical
enhancement or suppression of fluxes by cosmological selection
effects.  However, such models appear unnatural in the context of
string theory.  Hence, we expect that the true string theory landscape
will exhibit nontrivial cosmological selection effects.

\section{Discussion}
\label{sec-discussion}

We can understand the selection behavior qualitatively by studying the
tunneling rates, $e^{-B}$, where $B(q_i,n_i)$ is the instanton action
given in Eq.~(\ref{eq-B}).  Let us neglect tunneling events that
increase the magnitude of a flux $n_i$.  Then $\partial B/\partial
n<0$.  This means that among two unequal fluxes with similar charge
($q_i\approx q_j$), the larger flux, $\max\{|n_i|,|n_j|\}$, is more
likely to decay.  The dependence on the size of the charge is more
complicated.  For most of the parameter space explored here, one finds
$\partial B/\partial q>0$, which implies that among two equal fluxes
$|n_i|=|n_j|$, the flux associated with the smaller charge,
$\min\{q_i,q_j\}$, is more likely to decay.  (Exceptions can occur in
the final stages of the decay process.)

\begin{figure}
\begin{center}
\includegraphics[scale = .7]{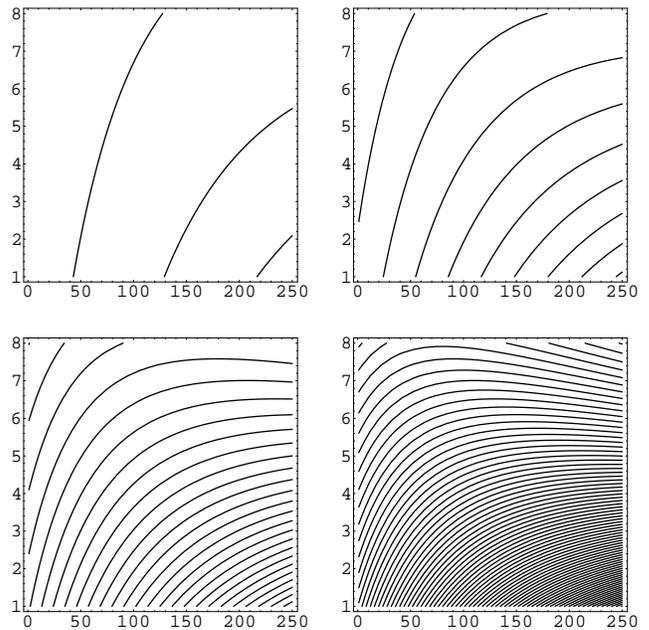}
\end{center}
\caption{Constant action surfaces for $\lambda=0.8$ (top-left),
  $\lambda=0.4$ (top-right), $\lambda=0.2$ (bottom-left), and
  $\lambda=0.1$ (bottom-right) in the $i$-$|n_i|$ plane.  (Recall that
  the charges $q_i$ are ordered so as to increase monotonically with
  $i$.)  The action $B$ grows towards the bottom right with a line
  spacing of $1$ in all four plots, corresponding to greater
  suppression of the decay.}
\label{fig-scheme}
\end{figure} 

In order to understand how these effects combine and compete, we have
plotted the lines of constant action in the flux-charge plane, for a
number of values of the cosmological constant $\lambda$, in
Fig.~\ref{fig-scheme}.  At the beginning of the decay chain, for large
$\lambda$, the action ranges only over an interval of a few, so the
most unlikely decay and the most likely decay do not differ enormously
in probability.  In this regime, our simulation shows that even upward
jumps (which are neglected in Fig.~\ref{fig-scheme}) are not very
suppressed.  

As $\lambda$ decreases below $0.8$, a preference for the decay of
fluxes with small $q_i$ and/or large $|n_i|$ becomes apparent.  The
slope of the constant action lines indicates how these tendencies are
balanced.  Starting around $\lambda=0.2$, the elimination of large
fluxes is strongly preferred.  This explains the absence of fluxes
$|n_i|\geq 5$ in our results (see Fig.~\ref{fig-pnis}, last panel).
In their absence, the last panel of Fig.~\ref{fig-scheme} shows a
strong preference for the decay of the fluxes associated with the
smallest few charges.  At least qualitatively, this explains the peaks
in the first few panels in Fig.~\ref{fig-pnis}.

\begin{figure}
\begin{center}
\includegraphics[scale = .75]{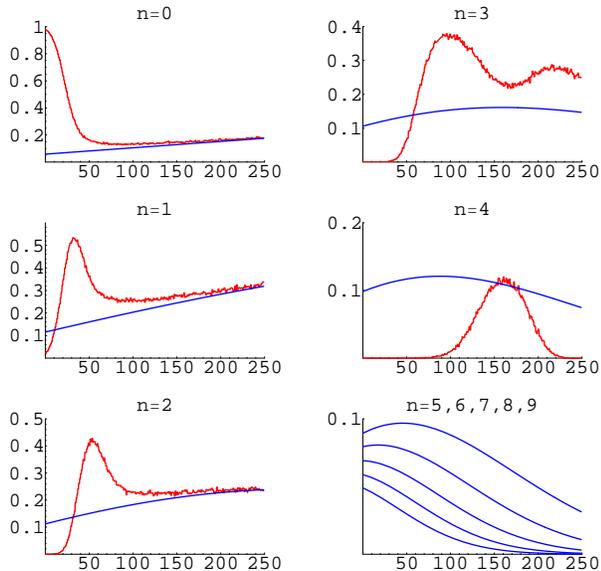}
\end{center}
\caption{The red, wiggly curves are the same as in Fig.~3.  The
  blue, smooth curves show the probabilities for various flux values
  near the beginning of the decay chain, at $\lambda=0.8$.  In the
  first few panels, it is apparent that small fluxes associated with
  large charges do not participate in the dynamics.  This can be
  understood from the functional form of the instanton action.}
\label{fig-012}
\end{figure}

Throughout the decay, in all four panels of Fig.~\ref{fig-scheme}, the
decay of fluxes with {\em both\/} small $|n_i|$ and large $q_i$ is
relatively suppressed.  This suggests that such fluxes will be left
untouched by the cosmological dynamics, and instead will be set by
initial conditions.  To check this conclusion, let us compare the
probability distribution $\tilde p_i(n)$ obtained from the $N$
penultimate vacua, with the distribution at the beginning of the decay
chain.  This is shown in Fig.~\ref{fig-012}.  The first three panels
($n\leq 2$) demonstrate that the initial probabilities are indeed
preserved for sufficiently large $i$.

\section{Comparison with an alternative measure}
\label{sec-discussion2}

We have computed the probability distribution $\tilde p_i(n)$ from the
measure proposed in Ref.~\cite{Bou06}.  Other cosmological measures
have been proposed, and it is important to ask which, if any, of the
proposals is on the right track.  In our view, this problem should be
treated no differently from any other challenge to come up with the
right theory describing a set of phenomena.  The measure problem, like
the question of the dynamical laws governing the early universe, is an
aspect of the theoretical challenge faced in quantum cosmology.  It
should be approached by a combination of seeking out simple guiding
principles (such as the causal-diamond viewpoint of the universe that
leads to the holographic measure used here), and elimination.  If a
measure, or a theory, is poorly motivated, then it is unlikely to get
much attention; and if a theory with a given measure yields absurd
predictions, then the theory or the measure or both must be wrong.

An interesting proposal which has been quite sharply formulated is the
measure of Garriga, Schwartz-Perlov, Vilenkin, and
Winitzki~\cite{GarSch05} (GSVW).  This measure has been applied to a
simplified BP model by Schwartz-Perlov and Vilenkin~\cite{SchVil06}
(SV), so let us compare results.

\subsection{Initial conditions and computational complexity}

Before going into the details, we note some formal differences.  The
GSVW measure depends on initial conditions only through the sign of
the initial cosmological constant.\footnote{More precisely, the
  measure is not defined for initial states composed exclusively of
  terminal vacua, and depends sensitively on initial conditions for
  initial states involving metastable $\lambda\leq 0$ vacua but no
  $\lambda>0$ vacua.}  We know of no fundamental reason why
cosmological dynamics should be required to have this property.  It
would be nice if we were spared the task of understanding the initial
conditions.  But even in the GSVW measure one would still need to
understand why the universe started in a de~Sitter vacuum.  

The measure we have used depends on initial conditions in additional
ways, for example through the value of the initial cosmological
constant.  As we have seen, however, the dependence is weak if we
assume that it enters mainly through $\lambda$.  But clearly, initial
conditions remain an free variable when applying the holographic
measure; we have argued only that within a wide range that appears to
be physically reasonable, the measure is not very sensitive to them.

It is unclear whether in practice, the GSVW measure can actually be
applied to a rich landscape such as that of string theory, as we have
done here for the holographic measure.  As shown in
Ref.~\cite{SchVil06}, the GSVW probability distribution is dominated
by the offspring of the longest-lived metastable vacuum.  This
requires finding the eigenvector $\mathbf{s}$ with smallest-magnitude
negative eigenvalue of a matrix with rank ten to the hundreds.
Perhaps statistical methods can be developed for this purpose, but so
far the problem is unsolved.  It is harder than the problem we
addressed here of statistically solving Eq.~(\ref{eq-matrix}), and it
may be computationally intractible in a realistic landscape, in the
sense of Ref.~\cite{DenDou06}.

In any case, SV considered only a toy-BP-model with $J=7$ charges and
a few million vacua.  Then the eigenvector $\mathbf{s}$ can be found
perturbatively in the small upward tunneling rates.  Because of the
much smaller number of vacua (in particular, their model does not
contain any vacua with $\lambda\sim\lambda_0$), we cannot directly
compare the results of SV to our results for a landscape with $\sim
10^{250}$ vacua.  

Even at a qualitative level, Ref.~\cite{SchVil06} contains no
predictions of likely flux distributions that could be compared to
ours.  However, there is one robust prediction of the GSVW measure,
which is not shared by the holographic measure, and which appears to
conflict with observation.  We turn to this problem next.

\subsection{The staggering problem}

The most important feature in the results of SV is the ``staggered''
nature of the probability distribution.  The GSVW probabilities are
widely spread.  At first order in the perturbative approximation
pursued by SV, the logarithmic range of probabilities, $\log \tilde
p_A$, appears to be larger than the number of vacua.  (A similar
conclusion was reached in Ref.~\cite{Sch06} for a different
landscape.)

This behavior is in sharp contrast to our distribution, and it is
extremely problematic.  As SV point out, if the staggered behavior
persisted at all orders in the full string landscape, the cosmological
constant problem could no longer be solved.  Then either the landscape
or the GSVW measure would be ruled out.\footnote{Until we discover
  alternatives to the string landscape in fundamental theory, or at
  least until we have exhaustively explored alternative measures, the
  more conservative conclusion would be to regard this problem as
  evidence against the GSVW measure, not against the string landscape.
  This is all the more true since in the holographic measure we have a
  well-motivated alternative that does not lead to this problem.}  At
first order in the SV approximation, staggering is generic, but one
might hope that at higher order, a more uniform probability
distribution will emerge~\cite{Sch06}.

However, we will argue that staggering is a very general feature of
the GSVW measure.  We will use a result proven in
Appendix~\ref{sec-mimic}: The GSVW measure can be perfectly mimicked
by the holographic measure, if the initial state $\mathbf{P}^{(0)}$ is
chosen to be the GSVW eigenvector $\mathbf{s}$, which is dominated by
the longest-lived metastable vacuum in the landscape, $(*)$.  This
result allows us to estimate the Shannon entropy of the GSVW
probability distribution by following the branching trees of
Ref.~\cite{Bou06}.

Unlike in the earlier sections, we will be estimating the probability
of vacua directly, not as a product of probabilities of individual
fluxes.  The Shannon entropy is then given by
\begin{equation}
  S_{\rm GSVW}= -\sum_A \tilde p_A \log \tilde p_A~,
\end{equation}
and the effective number of selected vacua is ${\cal N}_{\rm
  GSVW}=\exp(S_{\rm GSVW})$.  The cosmological constant problem can
only be solved if ${\cal N}_{\rm GSVW}\gg 10^{120}$.  In a staggered
probability distribution, the most likely few vacua dominate the
probability and the entropy, so $S_{\rm GSVW}\sim O(1)$, which is far
too small.  We will now argue that this type of distribution results
from the GSVW measure quite generally.

As SV point out, the $(*)$ vacuum, by virtue of being long-lived, will
have a relatively small cosmological constant, and upward tunneling
will be enormously suppressed.  Let us denote by $\epsilon$ the total
branching ratio for upward tunneling.  By Eq.~(\ref{eq-upward}),
$\log\epsilon\sim -1/\lambda_*$, where $\lambda_*$ is the cosmological
constant of the $(*)$ vacuum, so that $-\log\epsilon\gg 1$.  After one
upward tunneling, the cosmological constant is larger, but still only
of order $\langle n_i\rangle q_i^2\sim O(q_i/\sqrt{J})$, which is less
than $10^{-2}$ in natural models, so upward tunneling remains
extremely suppressed.  Following SV, let us thus use upward tunneling
as a small expansion parameter.  At zeroth order, we consider only
downward tunneling.

SV showed that the eigenvector $\mathbf{s}$ is dominated by one
component, the $(*)$ vacuum, and that the $(*)$ vacuum will have a
small enough cosmological constant to access only terminal vacua by a
single downward decay.  The arguments leading to this conclusion
generalize to realistic models.  However, to show that the GSVW
entropy is small, we will only rely on the weaker statement that the
GSVW-equivalent initial conditions correspond to starting from a small
number of initial vacua whose downward decay leads to terminal vacua
after very few steps.

Let us imagine that the downward decay of any metastable vacuum
affects each flux with equal relative probability, $1/J$.  Moreover,
let us pretend that decay chains never merge; all vacua reached by
downward decays are different.  These assumptions vastly overestimate
the spread of probabilities and thus overestimate the GSVW entropy,
for two reasons.  First, we are at small cosmological constant and
some decay channels will be much more suppressed than others;
vanishing fluxes cannot decay downward at all.  Second, the effective
number of selected vacua will also be smaller since some vacua will be
reached by more than one decay chain.

For simplicity, we assume that the tree contains only metastable vacua
until the $D$-th step, at which point all vacua will be terminal.
Starting from the the $(*)$ vacuum, $D$ will most likely be 1, but
when starting a downward progression after the first upward jump, $D$
can be a few, so let us keep it general.

At zeroth order, we thus have $D$ generations of vacua.  The $d$-th
generation contains $J^{d}$ vacua with normalized probability
$p_{d,0}\sim 1/DJ^d$ (the second index refers to the zeroth order).
A bound on the zeroth order entropy is therefore
\begin{equation}
  S_{\rm GSVW,0}<-\frac{1}{D}\sum_{d=1}^D \log\frac{1}{DJ^d} = \log D +
  \frac{D+1}{2}\log J~,
\label{eq-szero}
\end{equation}
This is of order one for realistic values of $J\lesssim O(1000)$ and
$D\sim O(1)$.  

The number of states, ${\cal N}_{\rm GSVW,0}\sim DJ^{\frac{D+1}{2}}$,
is very small compared to $10^{120}$---far too small to include vacua
with $\lambda\sim\lambda_0$.  Typically, anthropic considerations
cannot help in such a case, since most observers will be rare
fluctuations in a vacuum with $\lambda\gg\lambda_0$.  Hence, the GSVW
measure is not capable of populating the landscape at leading order.

Before turning to higher order corrections, let us pause to ask
whether the holographic measure would have succeeded at this stage,
i.e., ``at leading order'', when upward tunneling is neglected.
Indeed, our simulation obtains large entropy even if upward tunneling
is turned off.  However, one might be concerned that this is mainly
the result of starting from a large ensemble of vacua with similar,
large $\lambda$, rather than just a single vacuum (natural though such
a starting distribution may be~\cite{Vil02}).

But in fact the holographic measure would produce large $S$ even if we
restricted $\mathbf P^{(0)}$ to a single member of the initial
ensemble.  The only thing that matters is that we start from a generic
initial vacuum, for which $\lambda$ is of order unity.  Then there
will be a large number $D\sim O(\sqrt{J}/q)$ of downward tunnelings
($D\sim O(100)$ in our model) before the cascade terminates, and
Eq.~(\ref{eq-szero}) yields ${\cal N}\sim J^{O(\sqrt{J}/q)}$, which is
much larger than $10^{120}$ in generic models.

Returning to the GSVW measure, the entropy in Eq.~(\ref{eq-szero})
receives corrections at higher order.  We will now argue that they are
suppressed by powers of $\epsilon\ll 1$ and so they will hardly change
the zeroth order entropy.  In particular, they will not change the
conclusion that ${\cal N}_{\rm GSVW}\ll 10^{120}$.

At each order, we make one upward jump and gain one
``supergeneration'', or downward cascade of vacua, to which the above
analysis can again be applied.  Because upward tunneling is so highly
suppressed, it will almost certainly originate from the
largest-$\lambda$ vacuum in the $u$-th supergeneration and lead to a
strongly preferred vacuum with larger $\lambda$.  Corrections to this
assumption will be considered below.

Then the GSVW entropy may again be (over-)estimated by the normalized
probability distribution
\begin{equation}
p(u,d)=\frac{1-\epsilon}{D} \frac{\epsilon^u}{J^d}~,
\end{equation}
for a vacuum in the $d$-th generation of the $u$-th supergeneration.
Thus the GSVW entropy is bounded by\footnote{Here we have included the
  $(*)$ vacuum ($u=d=0$) to keep the expression simple.  Strictly, it
  should be excluded, but this leaves the result essentially unchanged
  quantitatively; an overall factor $(D+1)/D$ would appear on the left
  hand side of Eq.~(\ref{eq-asdf}).}
\begin{eqnarray}
  S_{\rm GSVW} &<& \sum_{u=0}^\infty \sum _{d=0}^D p(u,d)  \\
  & \approx & \log \frac{D}{1-\epsilon} + 
  \frac{\epsilon}{1-\epsilon} \log\frac{1}{\epsilon} +
  \frac{D+1}{2}\log J~. \nonumber
\end{eqnarray}
This differs from the zeroth order result mainly through the
negligible term
\begin{equation}
\epsilon \log \frac{1}{\epsilon}\ll 1~,
\label{eq-asdf}
\end{equation}
so the GSVW entropy remains too small even at higher order.

We have neglected the $u$-dependence of both $\epsilon\ll 1$ and $D$.
Choosing an upper bound $\epsilon\ll 1$ for all upward tunnelings and
$D\lesssim O(1)$ for the number of downward generations will only
overestimate the entropy but leave the conclusion unchanged.  But at
very high order, $u\sim O(\sqrt{J}/q)$, $\epsilon$ might be come
$O(1)$, $D$ will become larger, and the expansion might break down.
Can this change our conclusion?

To see that it will not, let us stop at $u$ of order a few, so that
our approximations hold up to this point.  Let us estimate the
remaining contributions to the entropy by imagining that at this point
the remaining probability flows equally into all vacua in the
landscape, resulting in ${\cal N}$ vacua with probability
$\epsilon^u/{\cal N}$.  (This is almost certainly a large overestimate
unless the landscape has no sinks or is extended far beyond the
semiclassical regime.)  But this contributes only of order
\begin{equation}
\epsilon^u \log \frac{\cal N}{\epsilon^u}
\end{equation}
to the GSVW entropy.  In a realistic landscape, we may safely assume
that $\log {\cal N} \ll \epsilon^{-u}$ even for $u\sim O(1)$.  Hence this
contribution is also negligible.  

We conclude that the GSVW entropy is dominated by the zeroth order
term, Eq.~(\ref{eq-szero}).  Its small size indicates that staggering
is a general problem for the GSVW measure.  It arises because the GSVW
measure, in effect, attempts to populate the landscape starting from
the longest-lived metastable vacuum.  This fails since it relies on
decay channels of enormously small relative probability.  With
reasonable initial conditions (such as starting with a randomly chosen
vacuum or an ensemble at large $\lambda$), this problem is absent in
the holographic measure.

\acknowledgments We thank Michael Douglas for discussions in which
some aspects of the approaches used in this paper first came up.  This
work was supported by the Berkeley Center for Theoretical Physics, by
a CAREER grant of the National Science Foundation, and by DOE grant
DE-AC03-76SF00098.

\appendix

\section{Statistical independence of fluxes}
\label{sec-independent}

Let us discuss the assumption that the $J$ fluxes are statistically
independent.  We will argue that there is some minor interdependence
because $J$ is finite, and we will explain how we correct for the
corresponding constraint.  Since the correction is small, some readers
may wish to skip this appendix.

If there are correlations between the distributions $p_i(n)$, then
their entropies do not simply add.  In this case, the total number of
states will be lower than the ``maximum randomness'' formula,
Eq.~(\ref{eq-ent}), would make us believe.  This will not be a concern
for our sample of $N$ initial vacua, nor for our unbiased samples of
vacua with small cosmological constant (without including selection
effects from cosmological dynamics).  As we discuss in
Appendix~\ref{sec-unbiased} below, one can generate suitable samples
by treating the fluxes as statistically independent.  However, it is a
concern for the $N$ vacua we arrive at by selection though a Monte
Carlo chain.

In fact, an obvious correlation is introduced by our choice to
consider the penultimate vacuum in each chain.  By construction, these
vacua must have a positive cosmological constant small enough that the
next flux decay can make $\lambda$ negative.  Thus, they satisfy the
constraint
\begin{equation}
2|\lambda_{\rm bare}|\leq \sum_i n_i^2
q_i^2<2|\lambda_{\rm bare}|+ (2n_j-1) q_j^2 
\label{eq-constraint}
\end{equation}
for at least one flux $j$.  This constraint would be unimportant if
$J$ was very large, but in realistic models, $J$ is only of order a
few hundred.

The number of vacua with distinct vacuum energies of order $\lambda_0$
was estimated in Eq.~(\ref{eq-nw}).  The correction due to the
constraint (\ref{eq-constraint}) can be approximately implemented as
follows.  The entropy of the unconstrained ensemble is larger than the
true entropy, with the extra states coming from vacua that fail to
satisfy Eq.~(\ref{eq-constraint}).  But consider the distribution of
vacua over the cosmological constant, $d\tilde{\cal N}/d\lambda$, in
both the constrained and unconstrained ensembles.  In the range of
$\lambda$ where Eq.~(\ref{eq-constraint}) is guaranteed to hold (which
ranges from $\lambda=0$ to an upper bound much larger than
$\lambda_0$), the two distributions $d\tilde{\cal N}/d\lambda$ will
agree well.  

From this viewpoint, the problem with Eq.~(\ref{eq-nw}) is not that
$\tilde {\cal N}$ is computed from the unconstrained ensemble; the
problem is that $dN/d\lambda$ is not.  But this is easy to correct.
Let us keep only the probabilities $\tilde p_i(n)$ obtained from the
Monte Carlo chain, but replace the actual sample of $N$ vacua
satisfying the constraint with a broader sample of $N'$ vacua randomly
generated from the $\tilde p_i(n)$, with the $\tilde p_i(n)$ treated
as truly independent.  This will spread the vacua over a larger range
of $\lambda$ and thus reduce the number of vacua with
$\lambda\sim\lambda_0$:
\begin{equation}
\tilde{\cal N}_{\lambda_0}=\lambda_0 \left.
\frac{d\tilde {\cal N}}{d\lambda}\right|_{\lambda=0}
\approx
\lambda_0 \frac{\tilde{\cal N}}{N'} \left.
\frac{dN'}{d\lambda}\right|_{\lambda=0}~,
\end{equation}
In the models studied in this paper, this correction reduces the
number of vacua with $\lambda\sim\lambda_0$ by about one order of
magnitude.

In principle, additional correlations could arise through the
dynamics.  By Eq.~(\ref{eq-B}), however, the decay rate of each flux
depends on the other fluxes only through $\lambda$ (and not, for
example, though terms like $(n_i-n_{i+1})^2$, which could introduce
clustering).  The vast majority of possible flux combinations at each
step in decay chains starting from a fixed value of $\lambda$ are
mapped to a small range of $\lambda$, so the relative decay rates
determining the next step cannot be sensitive to the detailed
distribution of fluxes.  

A very conservative upper bound on such correlations can be obtained
by ignoring averaging.  Instead, let us imagine that decay chains fall
into two families, one of which has higher values of $\lambda$ at each
step.  But they cannot differ by more than the amount by which the
final decay, at the end of the chain, changes $\lambda$ (or else, the
steps can just be relabeled so as to match the families more closely).
At the beginning of the chain, the decay is random for either family,
since $\lambda$ is large.  So the two families can differ at most by
the effect of one additional decay at the end.  Thus, we can bound
dynamical correlation effects by comparing our sample of penultimate
vacua with the ultimate vacua they decay to.  But in the models we
study, the corresponding probability distributions, $p_i(n)$, barely
differ.

\section{Generating an unbiased sample}
\label{sec-unbiased}

Consider all the vacua whose cosmological constant lies in a small
interval $\delta\lambda$ around some value $\lambda$.  On a number of
occasions in this paper, we require an unbiased sample of such vacua.
This task is equivalent to throwing all the vacua in the interval
$\delta\lambda$ into a bag, pulling one out, returning it\footnote{In
  the cases of interest here, the number of vacua in the interval
  $\delta\lambda$ will be much larger than the required sample size
  $\hat N$, so this step is not important.}, and repeating this
process $\hat N$ times to get a sample of size $\hat N$.

But the vacua are not in a bag; they are arranged in a grid on which
the above interval corresponds to a spherical shell (see
Fig.~\ref{fig-para}).  Because the grid has a high dimension $J$, we
must be careful to avoid directional effects that tend to pick vacua
from preferred patches of the shell.

To generate a sample, we first need a set of probability distributions
for the fluxes, $p_i(n)$.  Vacua can be generated by picking the
fluxes $n_i$ randomly according to the distributions $p_i(n)$.  It is
not necessary that all vacua thus generated lie in the interval
$\delta\lambda$.  We can reject those that do not, and generate vacua
until we have $\hat N$ that do.

However, this procedure will generate directional bias if the ensemble
of vacua defined by the $p_i(n)$ is not distributed spherically
symmetrically in flux space.  For example, if we specified that
$p_i(n)$ is identical for all $i$, then the distribution would not be
spherically symmetric since fluxes with large charge $q_i$ will tend
to contribute more to the cosmological constant.

A spherically symmetric ensemble can be generated as a kind of
canonical ensemble.  Let us define a ``temperature'', $T=1/\beta$,
dual to the cosmological constant $\lambda$.  Each flux contributes an
``energy'' $n_i^2 q_i^2/2$, so its partition function is
\begin{equation}
z_i(\beta) = \sum_{n_i=-\infty}^{\infty}
\exp \left(-\beta \frac{n_i^2 q_i^2}{2}\right)~.
\end{equation}
Treating the fluxes as independent, we obtain a total partition
function
\begin{equation}
Z(\beta)=\prod_{i=1}^J z_i(\beta) = 
\sum_{n_1,\ldots,n_J} e^{-\beta \Delta\lambda}~,
\label{eq-ensemble}
\end{equation}
where
\begin{equation}
\Delta\lambda(n_1,\ldots,n_J)\equiv \sum_{i=1}^J\frac{n_i^2 q_i^2}{2}~.
\end{equation}
Recall that $\lambda=-|\lambda_{\rm bare}|+\Delta\lambda$ by
Eq.~(\ref{eq-BP}).  Since the probability for each vacuum, $Z^{-1}e^{-\beta
  \Delta\lambda}$, depends on the fluxes only through the cosmological
constant, the ensemble is manifestly direction-independent.  

The individual partition functions, $z_i$, can be evaluated
analytically for\footnote{This condition is not overwhelmingly
  satisfied in the models studied here, so we evaluated the partition
  function numerically.  This turns out to make only a small
  difference, but all results in the main part paper were obtained in
  this more precise manner.  We use the analytic approximation in the
  formulas for ease of presentation.} $\beta q_i^2/2\ll 1$:
\begin{equation}
z_i(\beta)\approx q_i^{-1}\sqrt{\frac{2\pi}{\beta}}~,
\end{equation}
and the expectation value of the ``energy'' $\Delta\lambda$ is given
by
\begin{equation}
\langle\Delta\lambda\rangle(\beta)=\frac{J}{2\beta}~.
\end{equation}
In order to generate a sample containing vacua near $\lambda$, we
choose\footnote{It is not essential that $\beta$ is chosen just right,
  since in the cases of interest in this paper the ensemble
  (\ref{eq-ensemble}) has a larger spread than the interval
  $\delta\lambda$ that we want our sample to lie in.  As long as the
  distribution has significant overlap with it, a sample can be
  efficiently obtained by rejecting vacua that fail to lie in the
  desired interval.}
\begin{equation}
\beta=\frac{J}{2(\lambda+|\lambda_{\rm bare}|)}~,
\label{eq-beta}
\end{equation}
and pick the fluxes $n_i$ randomly according to the probability
distributions
\begin{equation}
p_i(n)=z_i^{-1}\exp \left(-\beta \frac{n^2 q_i^2}{2}\right)~.
\label{eq-pin}
\end{equation}

In this paper we have two distinct occasions to apply this method.
The first is the task of generating initial vacua for our Monte Carlo
chains.  For a given initial $\lambda$ we generate $N=1000$ initial
vacua according to Eqs.~(\ref{eq-beta}) and (\ref{eq-pin}).  From
them, we generate $1000$ decay chains and inspect the final
probability distribution $p_i(n)$ obtained from the penultimate vacua
on the chains.  We increase the initial $\lambda$ and repeat the
process until the final distribution shows asymptotic behavior, i.e.,
until an increase in the initial cosmological constant no longer
modifies our result.

The second application arises in estimating how many vacua with
$0<\lambda\leq\lambda_0$ are present in a given BP model, before
cosmological selection is taken into account.  We will begin by
discussing why the estimate given in Ref.~\cite{BP} is too crude in
the present context.

By Eq.~(\ref{eq-BP}), vacua with small cosmological constant
$0<\lambda\leq \lambda_0$ lie in a thin shell of radius
\begin{equation}
r=|2\lambda_{\rm bare}|^{1/2}
\end{equation}
and width
\begin{equation}
\Delta r=\lambda_0/|2 \lambda_{\rm bare}|^{1/2}
\end{equation}
in flux space.  Its volume is $\omega_{J-1}r^{J-1}\Delta r$, where
$\omega_{J-1}= 2\pi^{J/2}/\Gamma(J/2)$ is the area of a unit $J-1$
sphere.

In the limit where $r^2\gg \sum q_i^2$, directional effects in the
lattice can be neglected, and the number of vacua in the shell can be
estimated by comparing the volume of the shell to the volume of a
single cell in the grid, $\prod_{i=1}^J q_i$~\cite{BP}.  In the same
limit, the expectation values of the fluxes, $\langle n_i\rangle$,
will be much larger than unity.  Under the stronger condition that
$r^2\gg (\sum q_i)^2 J$, the probability that at least one flux
vanishes is very small.  Then most values of $\lambda$ in the interval
will be $2^J$-fold degenerate, and the number of different values of
$\lambda$ in the interval $0<\lambda<\lambda_0$ can be estimated as
\begin{equation}
\tilde {\cal N}_{\lambda_0}\approx 
\frac{|\pi\lambda_{\rm bare}/2|^{J/2}}{\Gamma\left(\frac{J}{2}\right)}
\, \frac{1}{\prod_{i=1}^J q_i}\, \frac{\lambda_0}{\lambda_{\rm bare}}
\label{eq-bpest}
\end{equation}

However, here we will study models in which the shell radii of
interest are of the same order of magnitude as the diagonal of a grid
cell: $r^2\sim\sum q_i^2$.  For the same reason, typical vacua on the
shell will have many vanishing fluxes ($n_i=0$).  The
$\lambda$-degeneracy of typical vacua will vary but will be less than
$2^J$.  Hence, Eq.~(\ref{eq-bpest}) is unreliable.

The canonical ensemble yields a better estimate.  With $\beta$ chosen
so that $\langle\Delta\lambda\rangle\approx |\lambda_{\rm bare}|$, we
obtain a canonical ensemble characterized by the probability
distributions (\ref{eq-pin}).  In order to account for degeneracies,
we treat vacua with the same cosmological constant as identical.  The
corresponding probability distributions $\tilde p_i(n)$, with $n\geq
0$, can be computed from Eq.~(\ref{eq-ptilde}). The full number of
vacua in this ensemble, $\tilde {\cal N}$ is given by
Eq.~(\ref{eq-tilden}).  The number of vacua in the tiny interval
$0<\lambda<\lambda_0$ is given by Eq.~(\ref{eq-nw}).  The distribution
$dN/d\lambda$ can be obtained by generating $N$ vacua randomly from
$p_i(n)$ [or $\tilde p_i(n)$] and binning them suitably.

\section{GSVW-equivalent initial conditions}
\label{sec-mimic}

In this appendix we show that our measure reproduces the GSVW
measure~\cite{GarSch05} if the initial probability distribution in
Eq.~(\ref{eq-matrix}) is taken to be a particular eigenvector singled
out by the GSVW analysis.

We should warn that this initial condition is a very unnatural choice
from the viewpoint of the holographic measure, and our result should
not be interpreted as an equivalence of the two measures in any
natural, physical sense.  Indeed, finding the appropriate eigenvector
$\mathbf{s}$ is the most challenging task in applying the GSVW
measure.  We are in effect using a nontrivial intermediate result of
the GSVW analysis as an initial condition for the holographic measure.
But this unified viewpoint allows us to reduce the difference between
measures to a difference in initial conditions.  In
Sec.~\ref{sec-discussion2}, we use this viewpoint to clarify both the
origin of ``staggering'' problem in the GSVW measure, and its absence
in our measure.

Let us briefly summmarize the GSVW measure.  It is derived from the
evolution of a ``comoving volume fraction'',
\begin{equation}
\frac{df_B}{dt}=\sum_A(-\kappa_{AB}f_B+\kappa_{BA}f_A)~,
\end{equation}
where $\kappa_{AB}$ is the rate per unit time for a worldline in
vacuum $B$ to enter vacuum $A$ (we neglect factors of the Hubble
parameter).  Let us restrict our attention to non-terminal vacua
($\lambda>0$), from whose volume fraction the GSVW probability
distribution is ultimately computed (see below).  We may then rewrite
the above equation using the fact that terminal vacua do not tunnel
back:
\begin{equation} 
\frac{df_b}{dt} = -\sum_A\kappa_{Ab}f_b+\sum_a\kappa_{ba}f_a~,
\label{eq-evo}
\end{equation}
where $a,b$ only run over non-terminal vacua.  

In the presence of terminal vacua, the solution to Eq.~(\ref{eq-evo})
is of the form
\begin{equation}
f_a(t)=s^{(1)}_ae^{-k_1t}+s^{(2)}_ae^{-k_2t}+\ldots
\label{eq-soln}
\end{equation} 
where $0<k_1<k_2\cdots$~\cite{GarSch05}.  The GSVW probability is
defined to be proportional to the ``bubble abundance'' produced by the
slowest decaying state, the vector $\mathbf{s}^{(1)}$:
\begin{equation}
p_A=\sum_b\kappa_{Ab}s_b^{(1)}
\label{eq-prob}
\end{equation}
Since the subleading terms are irrelevant, we will drop the sub- and
superscripts and write $\mathbf{s}\equiv \mathbf{s}^{(1)}$ and
$k\equiv k_1$.

For simplicity, we first demonstrate the claimed equivalence for the
probability of non-terminal vacua.  Let us write the above equations
in matrix form:
\begin{equation}
\frac{d\mathbf{f}}{dt}=(\mathbf{K} - \mathbf{J})\mathbf{f} ~,
\end{equation}
where $J_{ab}=\delta_{ab}\sum_A\kappa_{Ab}$ is a diagonal matrix, and
$\mathbf{K}$ is the matrix with components $\kappa_{ab}$.
The eigenvector $\mathbf{s}$ satisfies
\begin{equation}
({\mathbf{K} - \mathbf{J}}){\mathbf{s}}=-k{\mathbf{s}}.
\label{eq-eigen}
\end{equation}
The unnormalized GSVW probabilities for non-terminal vacua are
\begin{equation}
{\bf p=Ks}~.
\label{eq-prob1}
\end{equation}

By Eq.~(\ref{eq-rel}), the absolute decay rates are related to the
branching ratios by
\begin{equation}
{\bf E J = K}~,
\label{eq-relation}
\end{equation}
where $\mathbf{E}$ is the matrix with components $\eta_{ab}$.  Combining the
above four equations, we obtain
\begin{eqnarray}
(1-\mathbf{E}){\bf p} &=& (1-\mathbf{E}){\bf Ks} \nonumber \\
&=& (1-\mathbf{E})({\bf J}-k){\bf s} \nonumber \\
&=& ({\bf J-K}-k+k\mathbf{E}){\bf s}=k\mathbf{E} {\bf s}~.
\end{eqnarray}
Comparison with Eq.~(\ref{eq-matrix}), with initial condition
$\mathbf{P}^{(0)}=\mathbf{s}$, reveals that the GSVW probability
agrees with the holographic probability up to an overall factor $k$
that will be absorbed by normalization:
\begin{equation}
\mathbf{p}\propto\mathbf{P}~~\mbox{for}~~~\mathbf{P}^{(0)}=\mathbf{s}~.
\label{eq-equiv}
\end{equation}

To include terminal vacua, we extend the above argument to vectors and
matrices containing terminal components.  A prime will denote the
non-square matrix of decay rates from terminal into terminal
vacua.  First, we note that the GSVW probabilities are independent of
any terminal components $\mathbf{s}'$ ascribed to the eigenvector:
\begin{equation} 
\mathbf{p} = 
\left(\begin{array}{cc}
  \mathbf{K} & 0 \\
  \mathbf{K'}& 0 \\
\end{array}\right)
\left(\begin{array}{cc}
  \mathbf{s} \\
  \mathbf{s}' \\
\end{array}\right)=
\left(\begin{array}{cc}
  (\mathbf{J}-k)\mathbf{s} \\
  \mathbf{K's} \\
\end{array}\right) \\
\end{equation}
Similarly, by Eq.~(\ref{eq-matrix}), the holographic measure does not
depend on the terminal components of the initial probability
distribution $\mathbf{P}^{(0)}$.  For convenience, we might as well
set $\mathbf{s}'=0$.  

The remaining question is whether the (final) probabilities for
terminal vacua agree.  From Eq.~(\ref{eq-rel}) we have
\begin{equation}
\left(\begin{array}{cc}
  \mathbf{E} & 0 \\
  \mathbf{E}'& 0 \\
\end{array}\right)
\left(\begin{array}{cc}
  \mathbf{J} & 0 \\
  0 & 0 \\
\end{array}\right)=
\left(\begin{array}{cc}
  \mathbf{E}\mathbf{J} & 0 \\
  \mathbf{E}'\mathbf{J}& 0 \\
\end{array}\right)=
\left(\begin{array}{cc}
  \mathbf{K} & 0 \\
  \mathbf{K'}& 0 \\
\end{array}\right)~,
\end{equation}
so that
\begin{eqnarray}
  (1-\eta)
  \mathbf{p}&=&
  \left(\begin{array}{cc}
      1-\mathbf{E} & 0      \\
      -\mathbf{E}'          & 1 \\
\end{array}\right)
\left(\begin{array}{cc}
  (\mathbf{J}-k)\mathbf{s} \\
  \mathbf{K's} \\
\end{array}\right) \nonumber \\
&=& 
\left(\begin{array}{cc}
  (1-\mathbf{E})(\mathbf{J}-k)\mathbf{s}  \\
  \left[-\mathbf{E}'(\mathbf{J}-k)+\mathbf{K'}\right]\mathbf{s} 
\end{array}\right) \nonumber \\
&=&
\left(\begin{array}{cc}
  k\mathbf{E}\mathbf{s}     \\
  k\mathbf{E}'\mathbf{s}    \\
\end{array}\right)=k
\left(\begin{array}{cc}
  \mathbf{E}   & 0      \\
  \mathbf{E}'  & 0      \\
\end{array}\right)
\left(\begin{array}{cc}
  \mathbf{s}    \\
  \mathbf{0} \\
\end{array}\right) \nonumber \\
& = & k \eta \mathbf{s} ~, 
\end{eqnarray}
so Eq.~(\ref{eq-equiv}) holds for both nonterminal and terminal
components of the probability distribution.

Note that our result does not contradict a crucial difference between
the GSVW measure and the holographic measure, namely that the former
depends on absolute lifetimes of vacua whereas the latter knows
nothing about them.  (The former is constructed from the full matrix
$\kappa$; the latter only from its normalized version, $\eta$.)  To
achieve equivalence, we had to impart the extra information contained
in $\kappa$ to the holographic measure through a very special choice
of initial condition, $\mathbf{s}$.  We emphasize again that from the
point of view of the latter measure, this choice would be completely
unmotivated; the equivalence is useful mainly for clarifying the
differences of the two measures (Sec.~\ref{sec-discussion2}).

\bibliographystyle{board}
\bibliography{all}

\end{document}